\documentclass{article}

\usepackage{cite}
\usepackage{graphicx}
\usepackage{dcolumn}

\begin{document}

\date{}
\title{Exceptional points in two simple textbook examples}
\author{Francisco M. Fern\'{a}ndez \thanks{%
E-mail: fernande@quimica.unlp.edu.ar} \\
INIFTA (CONICET, UNLP), Divisi\'on Qu\'imica Te\'orica\\
Blvd. 113 S/N, Sucursal 4, Casilla de Correo 16, 1900 La Plata, Argentina}
\maketitle

\begin{abstract}
We propose to introduce the concept of exceptional points in intermediate
courses on mathematics and classical mechanics by means of simple textbook
examples. The first one is an ordinary second-order differential equation
with constant coefficients. The second one is the well known damped harmonic
oscillator. They enable one to connect the occurrence of linearly dependent
exponential solutions with a defective matrix that cannot be diagonalized
but can be transformed into a Jordan canonical form.
\end{abstract}

\section{Introduction}

\label{sec:intro}

Exceptional points\cite{HS90,H00,HH01,H04,GRS07} are being intensely studied
because they appear to be suitable for the explanation of a wide variety of
physical phenomena. Exceptional points (also called defective points\cite
{MF80} or non-Hermitian degeneracies\cite{BO98}) appear, for example, in
eigenvalue equations that depend on an adjustable parameter. As this
parameter changes two eigenvalues approach each other and coalesce at an
exceptional point. This coalescence is different from ordinary degeneracy
because the two eigenvectors associated to the coalescing eigenvalues become
linearly dependent at the exceptional point.

Numerical calculations predict the occurrence of exceptional points for the
hydrogen atom in cross magnetic and electric fields\cite{CMW07} and it has
been suggested that such points may be useful for estimating parameters of
some atomic systems\cite{ASKM16}. The predicted chirality of exceptional
points has been verified in a variety of experiments\cite
{DGHHHRR01,DDGHHHR03,DMBKGLMRMR16,XMJH16}. A detailed discussion of this
experimental work is beyond the scope of this pedagogical communication.
Many other articles about exceptional points are collected in the reference
lists of those just mentioned.

The concept of exceptional point can be introduced by means of simple
examples in intermediate courses on mathematics, quantum and classical
mechanics. The purpose of this paper is the discussion of two examples that
in our opinion are particularly appealing because they appear in courses of
mathematics and classical physics at the undergraduate level. In section~\ref
{sec:ode} we show the occurrence of exceptional points in ordinary
differential equations with constant coefficients. Such problem is discussed
in most textbooks on mathematics and differential equations\cite{A67}. In
section~\ref{sec:Damped_HO} we discuss the well known problem of a damped
harmonic oscillator when the friction force is proportional to the velocity
of the oscillating particle\cite{S67}. This problem has already been chosen
by Heiss\cite{H16} in his comment on experiments about encircling
exceptional points\cite{DMBKGLMRMR16,XMJH16}. Here we enlarge upon this
problem that may be of great pedagogical value. Finally, in section~\ref
{sec:conclusions} we summarize the main results and draw conclusions.

\section{Simple mathematical problem}

\label{sec:ode}

In order to illustrate some of the features of an exceptional point we
consider the second-order ordinary differential equation
\begin{equation}
\mathcal{L}(y)(x)=y^{\prime \prime }(x)+a_{1}y^{\prime }(x)+a_{0}y(x)=0,
\label{eq:ode_gen}
\end{equation}
with constant real coefficients $a_{i}$. One commonly obtains the solutions
from the roots of the polynomial
\begin{equation}
e^{-\alpha x}\mathcal{L}(e^{\alpha x})(x)=\alpha ^{2}+a_{1}\alpha +a_{0}=0,
\label{eq:poly_gen}
\end{equation}
that in this case are given by
\begin{equation}
\alpha _{1,2}=-\frac{a_{1}}{2}\pm \frac{1}{2}\sqrt{a_{1}^{2}-4a_{0}}.
\label{eq:roots_gen}
\end{equation}
The general solution is then\cite{A67}
\begin{equation}
y(x)=c_{1}e^{\alpha _{1}x}+c_{2}e^{\alpha _{2}x},  \label{eq:sol_gen}
\end{equation}
provided that $\alpha _{1}\neq \alpha _{2}$. The constants $c_{1}$ and $c_{2}
$ are determined by the initial conditions; for example $y(0)=y_{0}$ and $%
y^{\prime }(0)=y_{1}$.

A problem arises when $a_{1}^{2}=4a_{0}$ because the two solutions are
linearly dependent ($\alpha _{1}=\alpha _{2}=\alpha $). In this case the
differential equation is given by
\begin{equation}
\mathcal{L}(y)=\left( \frac{d}{dx}-\alpha \right) ^{2}y,  \label{eq:ode_ep}
\end{equation}
so that
\begin{equation}
\mathcal{L}\left( e^{\xi x}\right) =(\xi -\alpha )^{2}e^{\xi x}.
\end{equation}
On taking into account that
\begin{equation}
\frac{d}{d\xi }\mathcal{L}\left( e^{\xi x}\right) =\mathcal{L}\left( xe^{\xi
x}\right) =2(\xi -\alpha )e^{\xi x}+x(\xi -\alpha )^{2}e^{\xi x},
\end{equation}
we appreciate that the two linearly independent solutions are $e^{\alpha x}$
and $xe^{\alpha x}$; that is to way
\begin{equation}
y(x)=c_{1}e^{\alpha x}+c_{2}xe^{\alpha x}.  \label{eq:sol_ode_ep}
\end{equation}

Let us now approach the problem from another point of view. If we define $%
u=y^{\prime }$ then we can rewrite the second-order differential equation as
a first-order one in matrix form
\begin{equation}
\frac{d}{dx}\left(
\begin{array}{l}
y \\
u
\end{array}
\right) =\mathbf{A}\left(
\begin{array}{l}
y \\
u
\end{array}
\right) ,\;\mathbf{A}=\left(
\begin{array}{cc}
0 & 1 \\
-a_{0} & -a_{1}
\end{array}
\right) .  \label{eq:mat_dif_eq}
\end{equation}
The eigenvalues of the matrix $\mathbf{A}$ are exactly $\alpha _{1}$ and $%
\alpha _{2}$ and the corresponding unnormalized eigenvectors can be written
as
\begin{equation}
\mathbf{v}_{1}=\left(
\begin{array}{l}
1 \\
\alpha _{1}
\end{array}
\right) ,\;\mathbf{v}_{2}=\left(
\begin{array}{l}
1 \\
\alpha _{2}
\end{array}
\right) .  \label{eq:v1,v2_gen}
\end{equation}

When $a_{0}<$ $a_{1}^{2}/4$ the two eigenvalues are real. As $a_{0}$
approaches $a_{1}^{2}/4$ they approach each other, coalesce at $%
a_{0}=a_{1}^{2}/4$ and become a pair of complex conjugate numbers for $a_{0}>
$ $a_{1}^{2}/4$. Therefore, there is an exceptional point at $%
a_{0}=a_{1}^{2}/4$ where the matrix $\mathbf{A}$ becomes
\begin{equation}
\mathbf{A}=\left(
\begin{array}{cc}
0 & 1 \\
-\alpha ^{2} & 2\alpha
\end{array}
\right) ,  \label{eq:mat_A_ep}
\end{equation}
because $a_{1}=-2\alpha $ and $a_{0}=\alpha ^{2}$. Equation (\ref
{eq:v1,v2_gen}) clearly shows that $\mathbf{v}_{1}\rightarrow \mathbf{v}_{2}$
as $\alpha _{1}\rightarrow \alpha _{2}$ so that one eigenvector is lost at
the exceptional point as outlined in the introduction.

It is therefore clear from the argument above that at the exceptional point
the matrix $\mathbf{A}$ has only one eigenvalue and just one eigenvector
\begin{equation}
\mathbf{Av}_{1}=\alpha \mathbf{v}_{1},\;\mathbf{v}_{1}=\left(
\begin{array}{l}
1 \\
\alpha
\end{array}
\right) .  \label{eq:eigenvector_v1}
\end{equation}
The matrix $\mathbf{A}$ is not normal \textbf{(}$AA^{T}\neq \mathbf{A}^{T}%
\mathbf{A}$, where $T$ stands for transpose) and at the exceptional point it
becomes defective because it does not have a complete basis set of
eigenvectors; consequently, it cannot be diagonalized.

Consider a second vector $\mathbf{u}_{2}$ that is a solution to the equation
(Jordan chain\cite{GRS07})
\begin{equation}
\left( \mathbf{A}-\alpha \mathbf{I}\right) \mathbf{u}_{2}=\mathbf{v}_{1};
\end{equation}
for example
\begin{equation}
\mathbf{u}_{2}=\left(
\begin{array}{l}
0 \\
1
\end{array}
\right) .  \label{eq:eigenvector_v2}
\end{equation}
The matrix
\begin{equation}
\mathbf{U}=\left( \mathbf{v}_{1}\mathbf{u}_{2}\right) =\left(
\begin{array}{cc}
1 & 0 \\
\alpha  & 1
\end{array}
\right) ,  \label{eq:mat_U_gen}
\end{equation}
enables us to transform $\mathbf{A}$ into a Jordan canonical form
\begin{equation}
\mathbf{U}^{-1}\mathbf{AU}=\left(
\begin{array}{cc}
\alpha  & 1 \\
0 & \alpha
\end{array}
\right) .  \label{eq:mat_Jordan_gen}
\end{equation}
It is not diagonal but has the eigenvalues in the diagonal. Defective
matrices cannot be diagonalized but they can be transformed into Jordan
matrices.

This simple mathematical problem shows the appearance of exceptional points
in ordinary differential equations. At the exceptional point the exponential
solutions become linearly dependent and we have to look for another type of
solution. In addition to it, the matrix associated to the differential
equation becomes defective and cannot be diagonalized but can instead be
transformed into a Jordan matrix.

In closing this section we mention that the results shown above can be
generalized to ordinary differential equations of order $n$ with constant
coefficients
\begin{equation}
L(y)(x)=\sum_{j=0}^{n}a_{j}y^{(n)}(x)=0,
\end{equation}
where $y^{(0)}(x)=y(x)$ and $y^{(j)}(x)=\frac{d}{dx}y^{(j-1)}$, $%
j=1,2,\ldots ,n$. In order to obtain a matrix representation for this
equation we simply consider the column vector with elements $y^{(j)}$, $%
j=0,2,\ldots ,n-1$. We do not treat the general problem here because the
greater number of independent parameters in the resulting $n\times n$ matrix
makes the discussion of the exceptional points less appealing for
pedagogical purposes.

\section{Damped harmonic oscillator}

\label{sec:Damped_HO}

In this section we discuss a simple physical model: the damped harmonic
oscillator. A particle of mass $m$ moves in one dimension under the effect
of a restoring force that follows the Hooke's law: $f_{R}=-kx$, where $x$ is
the displacement from equilibrium. We also assume that there is a damping
force $f_{D}=-\gamma v$ that is proportional to the particle velocity $v=%
\dot{x}$, where the point indicates derivative with respect to time. This
force is due to friction between the particle and the medium and the
positive proportionality constant $\gamma $ is known as damping coefficient%
\cite{S67}. The Newton's equation of motion for the particle is
\begin{equation}
m\ddot{x}=-kx-\gamma \dot{x}.  \label{eq:eq_mot}
\end{equation}
If we apply the results of section~\ref{sec:ode} with $a_{0}=k/m$ and $%
a_{1}=\gamma /m$ we have to describe the properties of the physical system
in terms of two independent parameters. It is far more convenient to
transform equation (\ref{eq:eq_mot}) into a dimensionless one with just one
independent parameter.

In order to simplify the discussion of the model we define the dimensionless
time $s=\omega _{0}t$, where $\omega _{0}=\sqrt{k/m}$ is the frequency of
the oscillator when there is no damping. The resulting equation
\begin{equation}
y^{\prime \prime }(s)+y(s)+\lambda y^{\prime }(s)=0,  \label{eq:eq-mot-dim}
\end{equation}
where $\lambda =\gamma /\left( m\omega _{0}\right) $ and $y(s)=x\left(
s/\omega _{0}\right) $, is a particular case of (\ref{eq:ode_gen}) with $%
a_{0}=\lambda $ and $a_{1}=1$. The roots of the polynomial are
\begin{equation}
\alpha _{1,2}=-\frac{\lambda }{2}\pm \frac{1}{2}\sqrt{\lambda ^{2}-4}.
\label{eq:roots_damp}
\end{equation}

When $\lambda <2$ ($\gamma <2m\omega _{0}$) the two roots are complex and
the particle oscillates with frequency $\frac{1}{2}\sqrt{4-\lambda ^{2}}<1$,
or $\frac{\omega _{0}}{2}\sqrt{4-\lambda ^{2}}<\omega _{0}$ (underdamped
motion). The amplitude of the oscillation decreases exponentially as $\exp
\left( -\frac{\lambda }{2}s\right) $ because of the loss of energy due to
friction. If $\lambda >2$ the two roots are real and negative and the
particle approaches the equilibrium position without any oscillation
(overdamped motion). The particular case $\lambda =2$ is called critical
damping and $y(s)=c_{1}e^{-s}+c_{2}se^{-s}$ as discussed in section~\ref
{sec:ode}. We appreciate that there is no oscillation as in the preceding
case. The behaviour of this idealized physical system is discussed in detail
in any introductory textbook on classical mechanics\cite{S67}; here we are
mainly interested in the appearance of an exceptional point.

It is obvious, as discussed in section~\ref{sec:ode}, that there is an
exceptional point at $\lambda =2$. If we write the dimensionless equation of
motion as
\begin{equation}
\frac{d}{ds}\left(
\begin{array}{l}
y \\
u
\end{array}
\right) =\mathbf{A}\left(
\begin{array}{l}
y \\
u
\end{array}
\right) ,\;\mathbf{A}=\left(
\begin{array}{cc}
0 & 1 \\
-1 & -\lambda
\end{array}
\right) ,
\end{equation}
then the resulting matrix $\mathbf{A}$ cannot be diagonalized when $\lambda
=2$. Following the procedure outlined in section~\ref{sec:ode} we can
transform $\mathbf{A}$ into a Jordan matrix
\begin{equation}
\mathbf{U}^{-1}\mathbf{AU=}\left(
\begin{array}{cc}
-1 & 1 \\
0 & -1
\end{array}
\right) ,\;\mathbf{U}=\left(
\begin{array}{cc}
1 & 0 \\
-1 & 1
\end{array}
\right)
\end{equation}

\begin{figure}[tbp]
\begin{center}
\includegraphics[width=9cm]{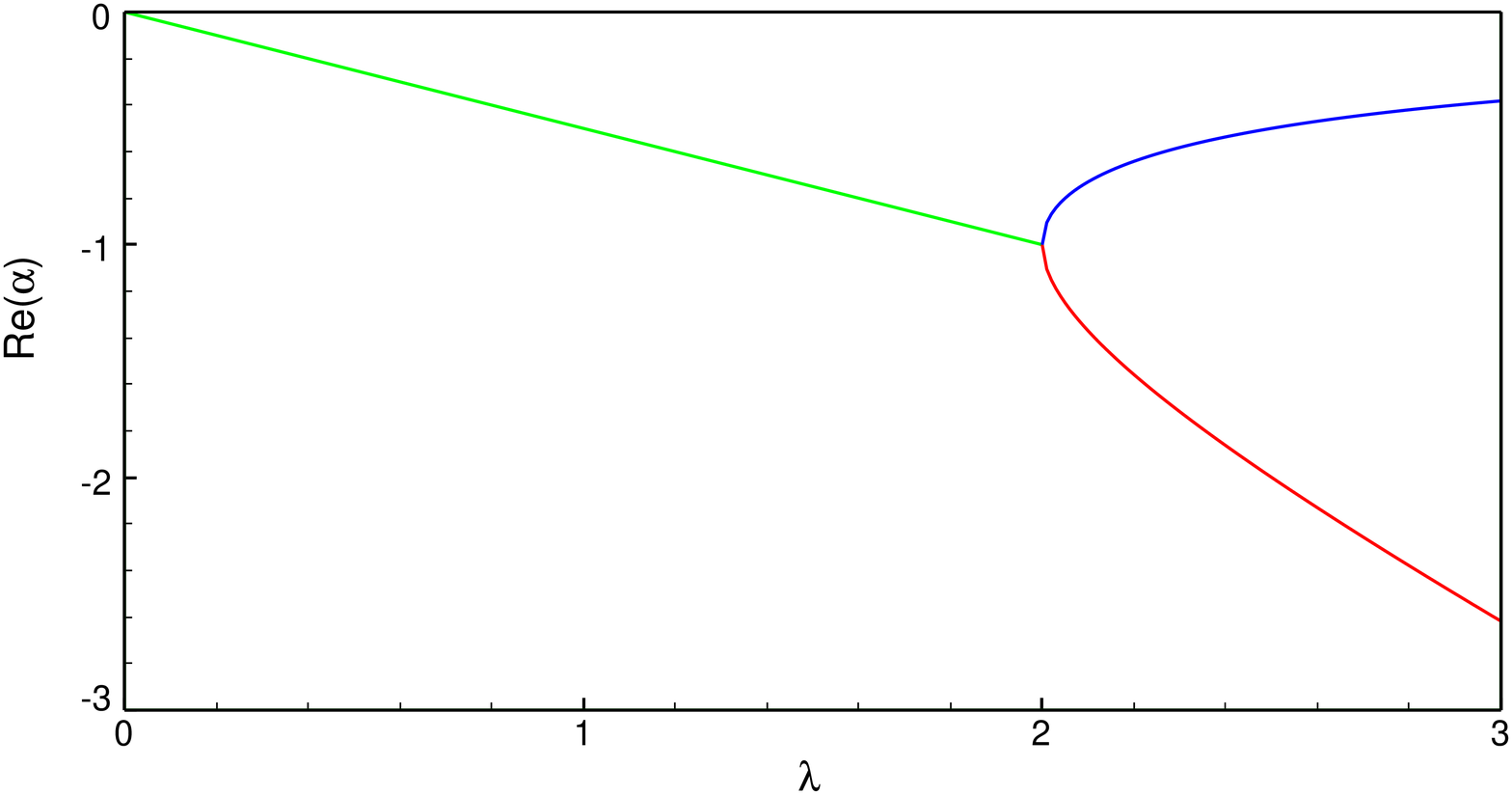} \includegraphics[width=9cm]{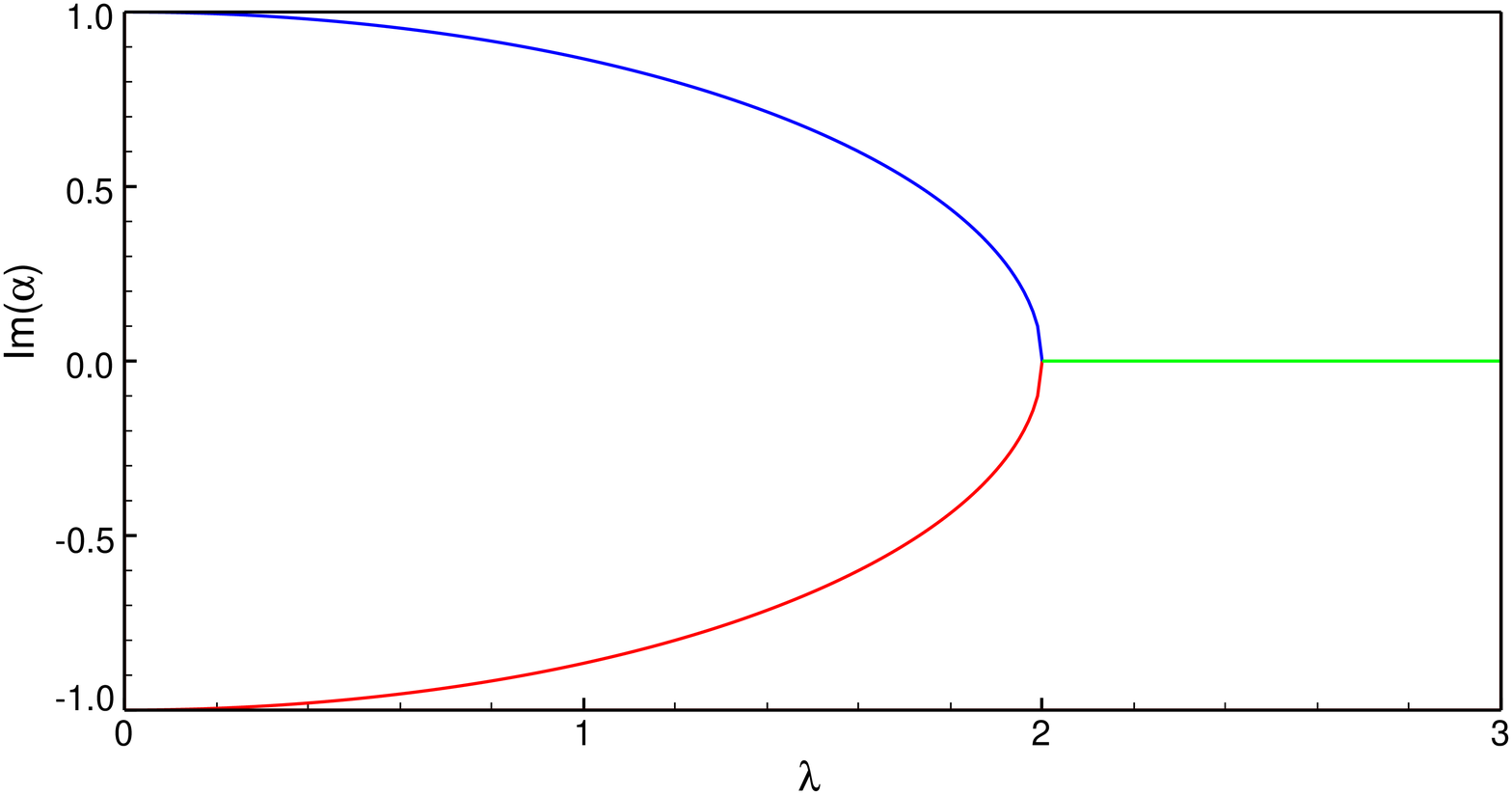}
\end{center}
\caption{Real and imaginary parts of $\alpha _{1}$ and $\alpha _{2}$ for the
damped harmonic oscillator}
\label{fig:alpha}
\end{figure}

Figure~\ref{fig:alpha}~shows the real and imaginary parts of the roots $%
\alpha _{1}$ and $\alpha _{2}$ for $0\leq \lambda \leq 3$. They exhibit the
well known branching pattern typical of exceptional points\cite{H04}. Note
that these figures apply to any combination of values of $m$, $k$ and $%
\gamma $ and that the use of a dimensionless equation enables us to describe
the main properties of the model in terms of only one independent parameter.

\section{Conclusions}

\label{sec:conclusions}

We think that the two examples discussed here are most suitable for the
introduction of the concept of exceptional point in intermediate courses on
mathematics and classical mechanics. The second-order ordinary differential
equation with constant coefficients is already discussed in almost any
textbook on mathematics or differential equations\cite{A67}. It shows the
connection between the occurrence of linearly-dependent exponential
solutions and the disappearance of one of the eigenvectors of the matrix
associated to the differential equation. The second example is one of the
simplest mechanical problems that exhibits friction and is also studied in
elementary courses on classical physics\cite{S67}. In this case the
exceptional point appears at the phase transition between underdamped and
overdamped motion (critical damping). It is worth noting that the use of a
dimensionless dynamical equation facilitates the analysis of the problem and
the plotting the eigenfrequencies for arbitrary values of the model
parameters: $m$, $k$ and $\gamma $. Instead of three parameters we need to
consider the variation of only one $\lambda =\gamma /\sqrt{mk}$ that
combines them in a suitable way.

\end{document}